\documentclass[prd,onecolumn,nofootinbib]{revtex4}
\pdfoutput=1
\usepackage{natbib}
\usepackage{longtable}
\usepackage{threeparttablex}
\usepackage{booktabs}    
\usepackage{graphicx}
\usepackage{tablefootnote}
\usepackage{amssymb,amsmath}
\usepackage{array}

\begin{document}

\title{A meta-analysis of impact factors of astrophysics journals}
\author{Rayani Venkat Sai Rithvik} \altaffiliation{E-mail: ee21btech11043@iith.ac.in}
\author{Shantanu Desai}
\altaffiliation{E-mail: shntn05@gmail.com}

\affiliation{$^{1}$Department of Electrical Engineering,
IIT Hyderabad,   Telangana-500084 India}

\affiliation{$^{2}$Department of Physics, IIT Hyderabad,   Telangana-500084 India}

\begin{abstract}
We calculate the 2024 impact factors for the  38  most widely used journals in Astrophysics, using the citations collated by NASA/ADS (Astrophysics Data System) and compare them to the official impact factors. This includes journals which publish papers outside of astrophysics such as PRD, EPJC, Nature, etc.  We also propose a new metric to gauge the impact factor based on the median number of citations in a journal and calculate the same for all the journals. We find that the ADS-based impact factors are mostly in agreement, albeit higher than the official impact factors for most journals. The journals with the maximum fractional difference in median-based and old impact factors are JHEAP and PTEP. We find the maximum difference between the  ADS and official impact factor for Nature.

\end{abstract}

\maketitle

\section{Introduction}
Every journal in the area of astrophysics has an associated impact factor. A few journals also have other metrics such as CiteScore, $h$-index, and Journal citation indicator. The Journal Impact Factor,  was originally
created as a tool to help librarians identify journals to purchase,
not as a measure of the scientific quality of research in an article~\citep{SFO}. However, it is one of the most widely used metrics to gauge the impact and importance of a journal. Note however that a number of caveats related to impact factors have been identified in astronomy literature~\citep{Frogel}.
In addition, the  citations in a journal do not always correlate with the the impact factor. For example, two very similar papers on the  abundance of Lithium in our galaxy have been published in Nature~\citep{SpiteNat} (impact factor of 50.5) as well as Astronomy and Astrophysics~\citep{SpiteAA} (5.4). However, among these, \citet{SpiteAA} has a larger  number of citations (898) compared to \citet{SpiteNat} (223). In this manuscript, we do a meta-analysis of the impact factors of some of the most widely used Astrophysics journals.

The  current method to calculate the official impact factor, which we refer to as \textbf{Old Impact Factor} in year \( n \) is defined as the ratio of the total number of citations in year \( n-1 \) of all papers published in the journal during years \( n-2 \) and \( n-3 \), divided by the number of refereed papers published in those same years~\citep{IF} \footnote{\url{https://en.wikipedia.org/wiki/Impact_factor}}

\begin{equation}
\text{IF}_{\text{old}}(n) = \frac{C_{n-1}}{P_{n-2} + P_{n-3}}
\end{equation}

where:
\begin{itemize}
    \item \( C_{n-1} \) is the total number of citations in year \( n-1 \) to papers published in years \( n-2 \) and \( n-3 \).
    \item \( P_{n-2} \) is the number of refereed papers published in the journal in year \( n-2 \).
    \item \( P_{n-3} \) is the number of refereed papers published in the journal in year \( n-3 \).
\end{itemize}
Therefore, the impact factor of a journal in 2024 is equal to the 
total citations received in 2023 for all papers published in that journal in  2021 and 2022. It does not include the citations published in the same year as the journal publication. The citations include both refereed and unrefereed publications. It is to be noted that sometimes an alternate definition has been used,  where the impact factor was defined as the average of citations in 2022 for papers published in 2021 and citations in 2023 for papers published in 2022~\citep{Abt}. It has also been noted that citation counts for astronomical papers peak at five years after publication~\citep{Abt81}. For this reason, it would also make sense to use a five year impact factor, which is sometimes reported for some journals. 

The official impact factors are calculated by the Clarivate company~\footnote{https://mjl.clarivate.com/home},   and are based on the citations  obtained using the  bibliometric data  from  Web of Science, which is owned by Clarivate. These citations are also sometimes  referred to as Science Citation Index (SCI)~\citep{Frogel}.

It has been pointed out that sometimes citations are missed or not attributed to the right person~\citep{Will}. Furthermore, errors in citations collated by the  Institute for Scientific Information  have been noted, because of non-standard conventions used by astronomers~\citep{Abt04}. Therefore, we re-evaluate the impact factors using citations collated by NASA/ADS~\citep{ADS}, which is the definitive resource and database for all astrophysics publications and compare them to the official impact factors.

Furthermore,  we also propose a new impact factor based on the  median number of citations, which we refer to as  \textbf{New Impact Factor}.
The \textbf{New Impact Factor} in year \( n \) is defined as the median of  citations in year \( n-1 \) to all the refereed papers published in the journal during years \( n-2 \) and \( n-3 \). This impact factor has also been previously defined in literature~\citep{Median1} and has been argued to be a better metric for cardiovascular journals~\citep{cardio}. However, this median-based impact factor has not been calculated for astrophysical journals. Sometimes, the impact factors of the  journals could show an abrupt rise due to a large number of citations in a given year~\citep{Abt},  and hence the new impact factor would be more robust to such fluctuations. 
However, we should note that both the mean and the median could be poor representations of the distribution of citations, which could be   generally very broad and dominated by its tails (in case of  power law distributions).  Despite this, an impact factor based on the mean could sometimes be more informative than one based on  median for some of the reasons mentioned below. Since the distribution of citations is a power law, papers with zero citations are typically the most represented ones in the distribution. There are many journals for which more than 50\% of the papers have no citations. All of these journals would be lumped into the same IF = 0 category, although they may have very different citation patterns. The median-based impact factor would not distinguish from a journal whose papers are never cited, and from a journal which has many citations but many papers with zero citations. With these caveats in mind, we now calculate the corresponding median-based impact factor for some of the most widely used  astrophysics journals and compare them to the usual way of calculating impact factor.

We have used NASA/ADS to obtain the number of citations for the calculation of the New Impact Factor.  For astrophysics, NASA/ADS is superior to most other databases and also reports a large number of citation metrics including tori index~\cite{tori}. This new impact factor would help assess the robustness of the impact factor. If there is a large difference, it would imply that the latest journal impact factors have been elevated because of only a handful of publications. 

We should point out that although some meta-analysis of citations and impact factors of a few astronomical journals have been done before~\citep{Abt,Abt04}, a large number of new journals in astrophysics have come up within the last two decades such as JCAP, Physics of Dark Universe, Open Journal of Astrophysics, Journal of High Energy Astrophysics, Astronomy and Computing, etc, which are now widely being used by astrophysicists for submitting manuscripts, because of their impact factors and no page charges. Among these, Open Journal of Astrophysics has not yet received an official impact factor at the time of writing. Furthermore, many Physics journals such as PRL, Physical
Review D, Physics Letters B, EPJC, PTEP, etc  are also regularly used for papers in some selected areas of Astrophysics, especially Cosmology, gravitational waves, and compact objects.  No such citation analysis for these new journals has previously been done in literature. Therefore, this is one of the motivations for doing such a study.

The manuscript is structured as follows. The methodology and results are described in Sect.~\ref{sec:results} and we conclude in Sect.~\ref{sec:conclusions}.

\section{Results}
\label{sec:results}
We considered 38 journals which are widely used in all areas of astronomy and astrophysics (including Cosmology, gravitational waves, and Particle Astrophysics). We have not considered journals in Planetary Science, Solar Terrestrial Physics and related areas, although it is straightforward to extend these studies to these or (any other) journals. 
We collated the citations using NASA/ADS API available at \url{https://ui.adsabs.harvard.edu/help/api/}. 
Some of the journals considered  such as PRD, PhRvL, PLB,  PTEP, EPJC, EL  also contain papers outside of Astrophysics, in the area of Particle Physics (EPJC, PRD), or in all areas of Physics (PhRvL, PLB, PTEP, EL). Some journals such as Nature and Science also  accept papers outside of Physics and Astronomy.
However, we only considered  astrophysics papers  published in the above journals which are tagged using ``collections:astronomy'' in ADS. 

\begin{figure}[h]
    \centering
    \includegraphics[width=\textwidth]{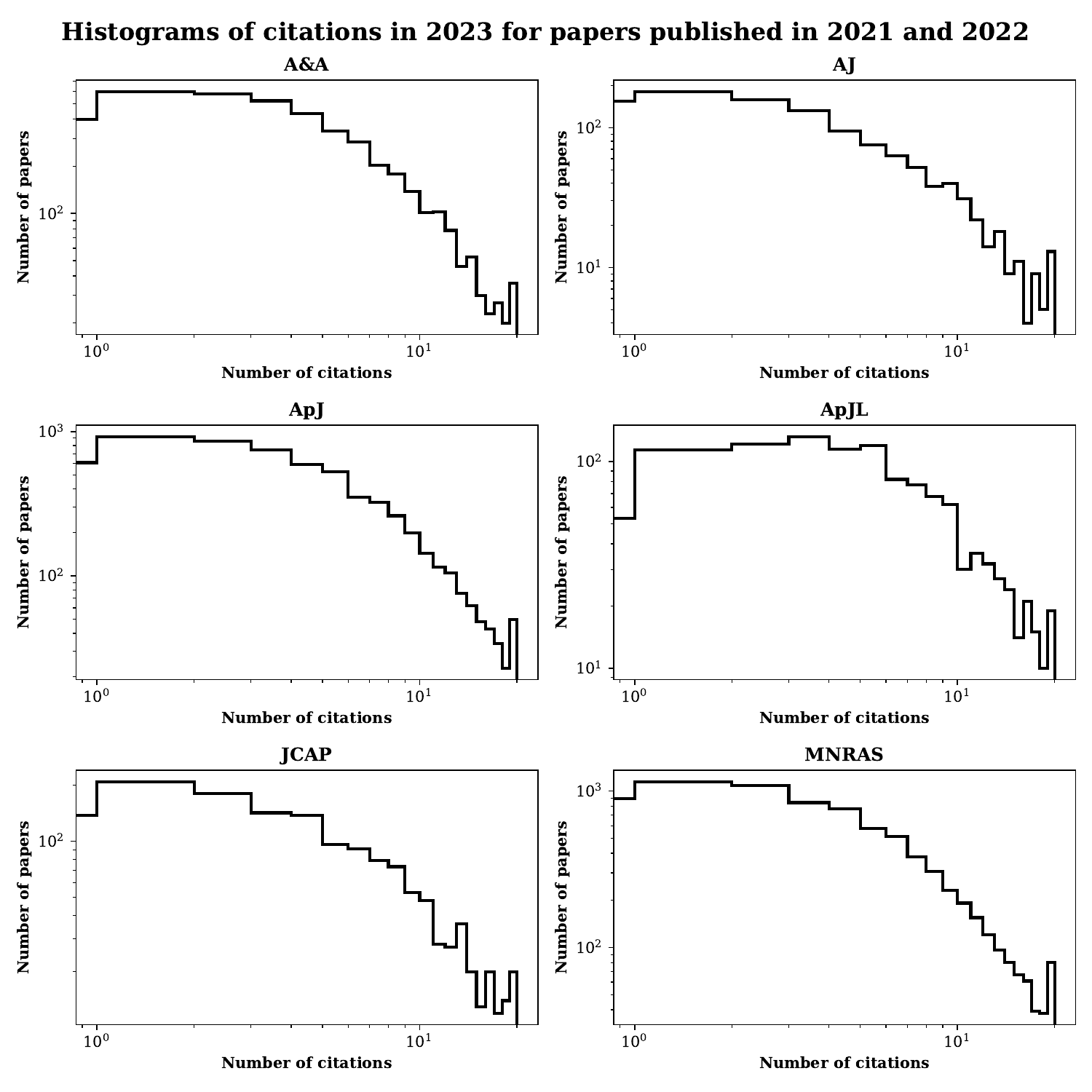}
    \caption{A histogram of number of citations (used in the calculation of 2024 impact factor) for six of the most widely used journals in Astrophysics using 20 logarithmically spaced bins. The citations show the power law distributions.}
    \label{fig:1}
\end{figure}

\begin{figure}[h]
    \centering
    \includegraphics[width=\textwidth]{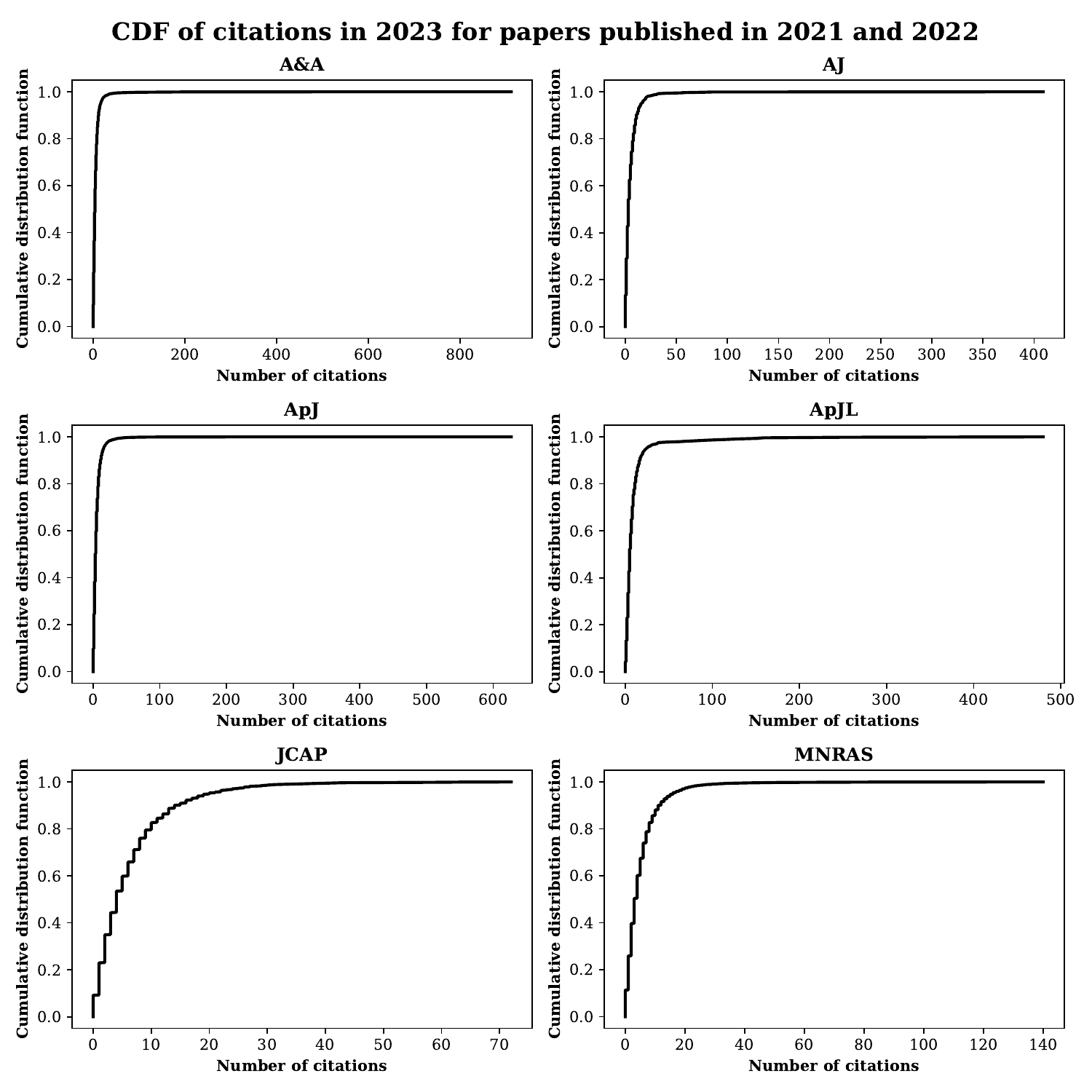}
    \caption{ Unbinned cumulative  distributions of the number of citations for the same six journals shown in Fig.~\ref{fig:1}.}
    \label{fig:2}
\end{figure}

Our results are summarized  in Table~\ref{tab1} with the full names of the journals in Table~\ref{tab2} in the Appendix.  For comparison, we have also shown the official impact factor available in the journal websites. Note that the Open Journal of Astrophysics (OJAP) does not yet have an official impact factor, as it has not yet been calculated  by Clarivate. Therefore, that column is left blank for OJAP. Furthermore, the official impact factor  for some of the Physics-based journals also  include non-astrophysics papers. We have also provided some additional publication and citation related diagnostics for each of these journals. These include the total number of published papers in 2021 and 2022, the fraction of papers without citations, and the number of citations of the most cited paper for each journal. All these metrics can be found 
in Table~\ref{tabnew}. We have also included the page charges or article processing charges (for subscription based access whenever available), and included those, based on the available data  as of Feb 2025\footnote{We note that this could be subject to change and most journals also offer waivers on request}. We have also plotted the histogram of the distribution of citations as well as the  unbinned cumulative citations for six of the most widely used journals in Astrophysics, viz. ApJ, AJ, A\&A, ApJS, MNRAS, and JCAP. These plots can be found in Fig.~\ref{fig:1} and Fig.~\ref{fig:2}.   
\clearpage
\begin{center}
\begin{longtable*}{|>{\centering\arraybackslash}p{3.5cm}|>{\centering\arraybackslash}p{2.0cm}|>{\centering\arraybackslash}p{3.0cm}|>
{\centering\arraybackslash}p{3.0cm}|}
\caption{Summary of old and new impact factors of 38 astrophysics journals for 2024. The first column indicates the journal abbreviation used by NASA/ADS. These abbreviations are defined in Table~\ref{tab2}. The second column shows the official impact factor indicated on the journal website. The third and fourth columns indicate the old and new (based on median) impact factors calculated using NASA/ADS. For all journals we  have only considered astrophysics publications in these journals which are tagged as ``collections:astronomy'' in NASA/ADS. However,  the official impact factor shown on the journal website also takes into account non-Astrophysics papers.}
\label{tab1} \\
\hline
\endfirsthead
\multicolumn{4}{c}
{{\bfseries \tablename\ \thetable{} -- continued}} \\
\hline
\endhead

\hline
\endfoot

\endlastfoot

  \textbf{Journal Code} & \textbf{Official Impact factor} & \textbf{(Old) Impact Factor using ADS} &  \textbf{New Impact Factor using ADS}    \\ 
  \hline
  A\&A & 5.4 & 6.00 & 4 \\ 
  \hline
  A \& A rev &27.8  &30.59 &13 \\ 
  \hline
   A\&C & 1.9 & 1.83 & 1 \\ 
  \hline
  AJ &  5.1 & 5.41 & 3 \\ 
  \hline
   AN & 1.1 & 1.04 & 0 \\ \hline
  ApJ & 4.8 &  5.37 & 3 \\ 
  \hline
   ApJL & 8.8 & 10.04 & 5 \\ 
  \hline
  ApJS & 8.6 & 9.30 & 4 \\ \hline
  Ap\&SS & 1.8 & 1.35 & 1 \\ 
  \hline
  APh & 4.2 & 4.37 & 1 \\ \hline 
  ARAA & 26.3 & 28.36 &21 \\ \hline
  AstL & 1.1 & 1.01 & 1 \\ \hline
  CQGra  & 3.6  & 4.22 & 2 \\ 
  \hline
   EL & 1.8 & 3.36 & 2 \\
  \hline
  EPJC & 4.2 & 4.63 & 3 \\ \hline
  EPJP & 2.8  & 2.82 & 1\\ \hline
  IJMPD & 1.8 & 2.33 & 1\\ 
  \hline
  Galax & 3.2 & 3.45 & 2 \\ \hline
   JApA & 1.1 & 1.17 & 0 \\ 
  \hline
  JHEAp  & 10.2 & 9.41 & 2 \\ \hline
  JCAP & 5.3 & 6.33 & 4 \\ 
  \hline
  MNRAS & 4.8 &  5.20 & 3 \\ 
  \hline
  Nat & 50.5 &8.49 &0 \\ \hline
  NatAst & 12.8 & 13.24 &2 \\ \hline
   NewA & 1.1 & 1.28 & 0 \\ \hline
   OJAP & N/A & 3.79 & 1 \\ 
  \hline
  PDU & 5.0 & 4.14 & 2 \\ 
  \hline
  PASA & 4.5 & 4.77 & 3 \\ 
  \hline
  PASJ & 2.2 & 2.59 & 1 \\ 
  \hline
  PASP & 3.3 & 3.51 & 1 \\ 
  \hline
  PHLB & 4.3 & 5.05 & 3 \\ 
  \hline
  PhRvL & 8.1 & 14.02 & 9 \\
  \hline 
  PRD & 4.6 & 6.43 & 4 \\  \hline
  PTEP  & 8.3  & 6.39 & 1 \\ 
  \hline
  RAA & 1.8 &  1.74 & 1 \\ \hline
  RvMP & 45.9 & 46.64 & 35 \\ \hline
  Sci & 44.7 &12.17 &1 \\ \hline 
  SSRV & 9.1  & 6.94 &5 \\ \hline 
  Univ &2.5 &  2.65 & 1 \\ \hline 
 \end{longtable*}
\end{center}

\begin{flushleft}
\begin{ThreePartTable}
\begin{TableNotes}
\footnotesize
\item [a] Page charges are waived if first author's affiliation is in a country that sponsors A \&A
\item [b] More details in \url{https://journals.aas.org/article-charges-and-copyright/}
\item [c] Page charges applicable only for articles submitted after 25th November 2024 
\item [d] No page charges if the primary classification of manuscript published on arXiv falls under hep-ph, hep-th, hep-ex, hep-lat
\end{TableNotes}
\begin{longtable*}{|>{\centering\arraybackslash}p{2.5cm}|>{\centering\arraybackslash}p{3.0cm}|>{\centering\arraybackslash}p{3.0cm}|>
{\centering\arraybackslash}p{3.0cm}|>
{\centering\arraybackslash}p{3.0cm}|}
\caption{ Some additional publication related statistics for  38 astrophysics journals for 2024 discussed in Table~\ref{tab1}. The page charges are based on available information as of Feb. 2025 and include charges for subscription based access if available and else refers to open access charges. Note also that ``-'' in the second column  indicates that the journal does not have any page charges for subscription access.  }
\label{tabnew} \\
\hline
\endfirsthead
\multicolumn{5}{l}
{{\bfseries \tablename\ \thetable{} -- continued}} \\
\hline
\endhead

\hline
\endfoot
\bottomrule
\insertTableNotes 
\endlastfoot
\textbf{Journal Code} & \textbf{Publication Charge} & \textbf{ \# Published (2021+2022)} &  \textbf{Fraction of Papers with no citations} & \textbf{Citations of top cited paper}    \\ 
  \hline
  A\&A & 100\texteuro/page\tnote{a} & 4325 & 0.1035 & 911\\ 
  \hline
  A \& A rev & -   & 17 & 0.0 & 101 \\ 
  \hline
   A\&C & -  & 127 & 0.3622 & 16 \\ 
  \hline
  AJ & $\geq$ \$1357\tnote{b}   & 1152 & 0.1441 & 409 \\ 
  \hline
   AN & -  & 238 & 0.5210 & 11 \\ \hline
  ApJ & Same as AJ &  6257 & 0.1082 & 627 \\ 
  \hline
   ApJL & $\geq$ \$2836 \tnote{b}   & 1255 & 0.0526 & 480 \\ 
  \hline
  ApJS & Same as AJ & 575 & 0.1165 & 261 \\ \hline
  Ap\&SS & - & 248 & 0.4395 & 28 \\ 
  \hline
  APh & - & 98 & 0.3469 & 105 \\ \hline 
  ARAA & - & 25 & 0.1200 & 83 \\ \hline
  AstL & - & 143 & 0.4266 & 7 \\ \hline
  CQGra  & -  & 802 & 0.2282 & 402 \\ 
  \hline
   EL & - & 69 & 0.2899 & 17 \\
  \hline
  EPJC & - & 841 & 0.1641 & 62 \\ \hline
  EPJP & -  & 234 & 0.2821 & 22\\ \hline
  IJMPD & -  & 289 & 0.3149 & 25\\ 
  \hline
  Galax & 1400 CHF & 240 & 0.2500 & 62 \\ \hline
   JApA & - & 212 & 0.5377 & 12 \\ 
  \hline
  JHEAp  & - & 56 & 0.3036 & 276 \\ \hline
  JCAP & - & 1507 & 0.0935 & 72 \\ 
  \hline
  MNRAS & 2310 \pounds &  7867 & 0.1256 & 140 \\ 
  \hline
  Nat & 9190 \pounds & 283 &0.9470 & 92 \\ \hline
  NatAst & 9190 \pounds & 292 &0.8527 & 63 \\ \hline
   NewA & - & 253 & 0.5099 & 21 \\ \hline
   OJAP & - & 34 & 0.3235 & 54 \\ 
  \hline
  PDU & - & 288 & 0.2257 & 31 \\ 
  \hline
  PASA & \$3550 \tnote{c} & 125 & 0.2400 & 61\\ 
  \hline
  PASJ & 6000\textyen /page & 243 & 0.2840 & 70 \\ 
  \hline
  PASP & \$3325 & 220 & 0.3318 & 156 \\ 
  \hline
  PHLB & - & 392 & 0.2041 & 53 \\ 
  \hline
  PhRvL & -  & 462 & 0.0368 & 231 \\
  \hline 
  PRD & - & 4352 & 0.0926 & 268 \\  \hline
  PTEP  & 130,000 \textyen \tnote{d}   & 88 & 0.2955 & 113 \\ 
  \hline
  RAA & - &  577 & 0.3328 & 40 \\ \hline
  RvMP & -  & 11 & 0.00 & 128 \\ \hline
  Sci & \$5450 & 109 & 0.4404 & 71\\ \hline 
  SSRV & -   & 163 & 0.1595 & 39 \\ \hline 
  Univ & \$2400 &  1198 & 0.3047 & 111 \\ \hline
\end{longtable*}
\end{ThreePartTable}

\end{flushleft}

Some of the salient features based on the results collated  in Table~\ref{tab1}, Table~\ref{tab2}, Fig.~\ref{fig:1} and  Fig.~\ref{fig:2} are as follows:
\begin{itemize}
\item Our results for the impact factor calculated using ADS agree with the official impact factor for almost all research journals within $\pm 1$.  The only exceptions are PRD, PTEP, PhRvL, ARAA, ApJL. The maximum difference is seen for 
PhRvL of about 6. However, the reason could be that  we have restricted our analysis to only Astrophysics-based publications, whereas the official impact factor includes publications outside of Astrophysics, which may not always be collated by ADS.

\item For most journals, the ADS impact factor is greater than the official one except for AP\&SS  (0.45),  A\&C (0.07), JHEAP  (0.8), PDU (0.86), PTEP  (1.9), RAA (0.06), AstL (0.09). This mostly agrees with previous such comparison studies. For example, ~\citet{Abtcitations} had noted that ADS has 15\% additional citations compared to SCI. ~\citet{Frogel}
had also pointed out  that the ratio of citations of ADS to SCI ranges from 1.22 to 2.17 between 2001 and 2006.

\item Most of the review papers (ARAA, A\&ARv, RvMP, SSRV) considered except RvMp have a difference between official and ADS based impact factors of greater than 1.
Among these review papers, A\&ARv has the largest difference between the new and old impact factor (of around 17).

\item The three journals with the maximum fractional difference between the new and old impact factor are   JHEAP (300\%) and  PTEP (500\%), where fractional increase is the ratio of the difference between the two impact factors divided by the new impact factor.  This is due to outliers in the number of citations for both  the papers, which we mention below.
For JHEAP this is due to the paper~\citet{Abdalla} which has 276 citations in 2023, while the second highest cited paper has 36 citations~\citep{Vagnozzi}. Both of these publications are in the area of Cosmology. For PTEP, there are three papers with citations over 50 which are ~\citet{KAGRA,DECIGO,Mei} having citations of 96, 82, and 60 respectively. All the three publications are related to gravitational wave detectors. 
\item If we consider journals with high impact factors, which include non-astrophysics journals such as Nature, Science, PRL, the maximum difference between the official and ADS-based impact factor is Nature with a difference of 42. The journal Science shows a corresponding difference of 32. Also the new impact factor of Nature is zero, as most of the astronomy related publications in Nature have zero citations in 2023. One possible reason for the low value of the impact factor for Nature  could be due to  the advent of the journal ``Nature Astronomy'', which is increasingly being used to supersede Nature for astrophysics related papers.  
Another  possibility could be due to the fact that many papers published by Nature/Science and indexed in NASA/ADS are comments (perspectives) rather than normal research papers. These commentary articles usually receive low or zero citations. It is not straightforward to segregate such commentary articles from normal research papers.~\footnote{We are thankful to Zhaozhou Li for pointing this out.}
This also reinforces the fact that Nature is not the best journal for Astrophysics. 
Having said that, we should point that this result could be a statistical fluke and applicable only for calculating the impact factor for 2024. 
We also note that in terms of publication charge Nature is one of the most expensive among all contemporary astrophysics journals. 

\item From Fig.~\ref{fig:1}, we see that the distribution of citations of the six most widely used journals follow a power-law.
This shows that median-based impact factor may not be the best measure for such journals.
None of the journals show a conspicuous peak in the distribution of citations.  Similarly the CDF of A\&A, AJ, ApJ and ApJL show a sharp rise, whereas the same for JCAP and MNRAS show a smooth rise.

\end{itemize}
\section{Conclusions}
\label{sec:conclusions}
In this work, we have done an extensive meta-analysis of citations for some of the most widely used Astrophysics journals, including new journals from the last two decades.
We have independently calculated the 2024  impact factors of 38 Astrophysics and Physics journals which accept astrophysics papers, using NASA/ADS database and compared them to the official impact factor of each journal, which have been obtained using the SCI based citations calculated by Clarivate.  We also proposed a new impact factor based on the median number of citations and calculated the same for all the journals.  Our results for all the three impact factors can be found in Table~\ref{tab1}. We have also provided additional publication and citation related diagnostics, including page charges for all these journals in Table~\ref{tabnew}. We have also shown a histogram of the number of citations and its CDF in Fig.~\ref{fig:1} and Fig.~\ref{fig:2}, respectively.

We find that the impact factors using ADS for most research journals are in agreement  with  the official impact factors. 
The maximum difference is obtained  for PhRvL (6). However, this maybe due to the fact that our analysis only considers astrophysics  journals, whereas the official impact factor also includes non-astrophysics journals.
However for most journals, the ADS based impact factor is higher than the official impact  factor.  This is due to the fact that the citations in ADS are larger than 
that in SCI, which has been noted before~\citep{Frogel}. 
The journals showing the largest fractional difference between new and old impact factors are JHEAP and PTEP of 300-500\%.  We also find the maximum difference between ADS and official impact factor for Nature of around 50. The new impact factor for Nature is 0. Therefore, the high impact factor of Nature is currently being driven by  publications outside of astrophysics and it may not be the most effective journal for astrophysics.

In a future work, we shall also do a similar analysis for the five year impact factors given the observations in ~\citet{Abt81}. We however note that the concept of impact factor is still fraught with caveats  and does not tell
us which journal is useful for research.
In the spirit of open science, we have made our analysis codes publicly available at \url{https://github.com/Rithvik-2003/ImpactFactor/}, which anyone can use to do a similar study for any other journal, whose papers are indexed in NASA/ADS.

\section*{Acknowledgments}

The authors thank NASA/ADS staff for prompt help during this analysis and answering all our queries. We are also grateful to Choong Ngeow,  Zhaozhou Li, and three  anonymous referees for useful feedback and suggestions on this manuscript.

\section*{Data availability}
There are no data associated with the article. The codes for analysis can be found on GitHub.
\section*{Appendix}

\begin{longtable*}
{|>{\centering\arraybackslash}p{3.0cm}|>{\centering\arraybackslash}p{9cm}|}
\caption{Journal Names and their corresponding Codes/Abbreviations.
\label{tab2}}
\\
\hline
\endfirsthead
\multicolumn{2}{c}%
{{\bfseries \tablename\ \thetable{} -- continued}} \\
\hline
\endhead

\hline
\endfoot
\endlastfoot

  \textbf{Journal Code} & \textbf{Journal Name}   \\ 
  \hline
  A\&A & Astronomy \& Astrophysics\\ 
  \hline
  A \& A rev & Astronomy and Astrophysics Review \\ \hline
   A\&C & Astronomy and Computing  \\ 
  \hline
  AJ & Astronomical Journal \\ \hline
   AN & Astronomische Nachrichten  \\ \hline
   APh & Astroparticle Physics \\ \hline
  ApJ & The Astrophysical Journal  \\ 
  \hline
   ApJL & The Astrophysical Journal Letters \\ 
  \hline
  ApJS & The Astrophysical Journal Supplement Series \\ \hline
  Ap\&SS & Astrophysics and Space Science \\ 
  \hline
  ARAA & Annual Review of Astronomy and Astrophysics  \\ \hline
  AstL & Astronomy Letters   \\ \hline
  CQGra  & Classical and Quantum Gravity  \\ 
  \hline
   EL & Europhysics Letters  \\
  \hline
  EPJC & The European Physical Journal C \\ \hline
  EPJP & European Physical Journal Plus \\ \hline
  Galax & Galaxies \\ \hline 
  IJMPD & International Journal of Modern Physics D \\ 
  \hline
   JApA & Journal of Astrophysics and Astronomy \\ 
  \hline
  JHEAp  & Journal of High Energy Astrophysics \\ \hline
  JCAP & Journal of Cosmology and Astroparticle Physics \\ 
  \hline
  MNRAS & Monthly Notices of the Royal Astronomical Society \\ 
  \hline
  Nat & Nature \\ \hline
  NatAst & Nature Astronomy \\ \hline
   NewA & New Astronomy \\ \hline
   OJAP & The Open Journal of Astrophysics  \\ 
  \hline
  PDU & Physics of the Dark Universe  \\ 
  \hline
  PASA & Publications of the Astronomical Society of Australia \\ 
  \hline
  PASJ & Publications of the Astronomical Society of Japan \\ 
  \hline
  PASP & Publications of the Astronomical Society of the Pacific  \\ 
  \hline
  PHLB & Physics Letters B \\ 
  \hline
  PhRvL & Physical Review Letters \\
  \hline 
  PRD & Physical Review D  \\  \hline
  PTEP  & Progress of Theoretical and Experimental Physics  \\ 
  \hline
  RAA & Research in Astronomy and Astrophysics  \\ \hline
  RvMP & Reviews of Modern Physics \\ \hline
  Sci & Science \\ \hline
  SSRV & Space Science Reviews \\ \hline
  Univ & Universe \\ 
  \hline

\end{longtable*}

\bibliography{Wiley-ASNA}%

\end{document}